\documentclass[prd,preprint,eqsecnum,nofootinbib,amsmath,amssymb,preprintnumbers,tightenlines,longbibliography,floatfix]{revtex4-1}

\usepackage[mathscr]{eucal}
\usepackage{mathrsfs}
\usepackage{hyperref}
\usepackage{graphicx}
\usepackage{bm}

\usepackage{graphicx}
\usepackage{bm}

\def\r{{\bm r}}

\def\st{\begin{equation}}
\def\stp{\end{equation}}
\def\bg{\begin{eqnarray}}
\def\nd{\end{eqnarray}}
\def\Eq#1{Eq.~(\ref{#1})}

\def\Fig#1{Fig.~\ref{#1}}
\def\Sect#1{Section~\ref{#1}}

\def\Eq#1{Eq.~(\ref{#1})}

\def\Fig#1{Fig.~\ref{#1}}
\def\Sect#1{Section~\ref{#1}}

\def\nott#1{\setbox0=\hbox{$#1$}                
   \dimen0=\wd0                                
   \setbox1=\hbox{/} \dimen1=\wd1               
   \ifdim\dimen0>\dimen1                        
      \rlap{\hbox to \dimen0{\hfil/\hfil}}      
      #1                                        
   \else                                        
      \rlap{\hbox to \dimen1{\hfil$#1$\hfil}}   
      /                                         
   \fi}                                         %

\advance\parskip 1.9pt
\advance\voffset -0.2in

\begin{document}
\title{Fusion Probability in Dinuclear System}
\author{Juhee Hong}
\affiliation{\mbox{Rare Isotope Science Project, 
Institute for Basic Science,} \\
Daejeon 305-811, Korea}
\date{\today}

\begin{abstract}
Fusion can be described by the time evolution of a dinuclear system with 
two degrees of freedom, the relative motion and transfer of nucleons. 
In the presence of the coupling between two collective modes, we solve the 
Fokker-Planck equation in a locally harmonic approximation. 
The potential of a dinuclear system has the quasifission barrier and the 
inner fusion barrier, and 
the escape rates can be calculated by the Kramers' model. 
To estimate the fusion probability, we calculate the quasifission rate and the 
fusion rate. 
We investigate the coupling effects on the fusion probability and the cross 
section of evaporation residue. 
\end{abstract}

\maketitle

\section{Introduction}

Fusion reaction has been greatly investigated, and superheavy elements have 
been synthesized in experiments. 
Many theoretical models describe heavy ion reaction and estimate the cross 
section of evaporation residue. 
Although much effort has been devoted to understanding fusion, it is still 
challenging to analyze and to predict data quantitatively\cite{aritomo}. 
Especially, the fusion probability is the least understood factor contributing 
to the cross section. 
The fusion probability has been studied as diffusion, and it is expected to 
be a logistic function\cite{FPkramers,othercite2,othercite}. 
However, it might be sensitive to details of dynamics and 
be important for more 
quantitative analysis of data. 
In this work, we investigate the fusion probability by using 
the Fokker-Planck equation in the dinuclear system concept. 

A dinuclear system consists of two colliding nuclei captured in an almost 
touching configuration\cite{dns}. 
In this concept, fusion can be understood as the time evolution of 
a dinuclear system. 
This description is different from other models in that it 
retains the individuality of nuclei and considers quasifission. 
A dinuclear system has two degrees of freedom, the relative motion and 
transfer of nucleons. 
To describe the relative motion, the internuclear 
distance $R$ is used as a collective coordinate. 
For transfer of nucleons, we define the mass asymmetry parameter
\st
\eta=\frac{A_1-A_2}{A_1+A_2} \, ,
\stp
where $A_1$ and $A_2$ are mass numbers of two nuclei. 
Each mode of motion is a diffusion process in a potential with an 
energy barrier. 
The escape rate over the barrier can be calculated by the Kramers' model. 
There has been much research on the Kramers' model (for a review and 
references, see Ref.\cite{kramers2}). 
The extension of the model can be applied to fusion reaction. 

The time evolution of a dinuclear system is described by the Fokker-Planck 
equation. 
For asymmetric reactions with not too high mass asymmetry, 
we have the following equation\cite{zpa,FPkramers}:
\bg
\label{dnseq}
\frac{\partial f}{\partial t}
&=&-\mu_{R}p_R\frac{\partial f}{\partial R} 
-\mu_{\eta}p_\eta\frac{\partial f}{\partial \eta} 
-\left(\frac{\partial \mu_{R}}{\partial \eta}\frac{\mu_{\eta}}
{\mu_{R}}p_\eta+\frac{\partial \mu_{\eta}}{\partial R}\frac{\mu_{R}}
{\mu_{\eta}}p_R\right)f \nonumber\\
&&+\left(\frac{\partial U}{\partial R}- \frac{\partial \mu_{R}}
{\partial \eta}\frac{\mu_{\eta}}{\mu_{R}}p_R p_\eta
+\frac{1}{2} \frac{\partial \mu_{\eta}}{\partial R}p_\eta^2
\right)\frac{\partial f}{\partial p_R}
+\left(\frac{\partial U}{\partial \eta}- \frac{\partial \mu_{\eta}}
{\partial R}\frac{\mu_{R}}{\mu_{\eta}}p_R p_\eta+
\frac{1}{2}\frac{\partial \mu_{R}}{\partial \eta} p_R^2
\right)\frac{\partial f}{\partial p_\eta}
\nonumber\\
&&+\gamma_R\mu_R\frac{\partial(p_R f)}{\partial p_R}
+\gamma_\eta\mu_\eta \frac{\partial(p_\eta f)}{\partial p_\eta}
+D_R\frac{\partial^2f}{\partial p_R^2}
+D_\eta\frac{\partial^2f}{\partial p_\eta^2} \, ,
\nd
where $\mu_i$, $\gamma_i$, and $D_i$ ($i=R, \, \eta$) are the diagonal 
components of the inverse mass tensor, the friction tensor, and the diffusion 
tensor, respectively. 
By the fluctuation-dissipation theorem, $D_i=\gamma_i T_i$, where 
$T_i=(\hbar\omega_i/2) \, \mbox{coth}(\hbar\omega_i/2T)$ is the effective temperature 
with the energy of zero oscillations $\hbar\omega_i/2$ and the thermodynamic 
temperature $T=\sqrt{E^*/a}$ ($a=A/12 \, \mbox{MeV}^{-1}$ is the level density 
parameter). 

The Kramers' problem described by \Eq{dnseq} is useful in other quantum 
mechanical systems as well as a dinuclear system. 
The escape rate in two-dimensional potential has been 
calculated\cite{langer,2drate}, but we use a different strategy to calculate 
the diffusion rates of two collective modes. 
By assuming that two modes are weakly coupled through the 
inverse mass tensors, we solve the Fokker-Planck equation perturbatively. 
We then calculate the quasifission rate and the fusion rate to estimate the 
fusion probability. 
Although we apply the rates to quasifission and fusion, our results for the 
diffusion rates can be applied to any escape problem in the presence of 
friction and diffusion. 

In \Sect{kramersbasic}, we review the Kramers' model to calculate the 
escape rate. 
In \Sect{pcncorrection}, we extend the model with two collective modes which 
are coupled through the inverse mass tensors. 
For weak coupling, we solve \Eq{dnseq} perturbatively and calculate the 
diffusion rates of two modes. 
By applying the escape problem to a dinuclear system, we consider 
asymmetric fusion reactions in \Sect{dnsreview}. 
We review the dinuclear system concept and estimate the coupling effects on 
the fusion probability and the cross section of evaporation residue. 
Finally, we summarize our results in \Sect{sumdiscuss}.

\section{Kramers' Model}
\label{kramersbasic}

The fusion probability can be estimated by the Kramers' model. 
Kramers considered the escape problem as a one-dimensional Brownian motion in 
a potential with deformation energy\cite{kramers}. 
In this section, we review the Kramers' model, and the extension with two 
degrees of freedom will be discussed in the next section. 

\begin{figure}
\includegraphics[width=0.45\textwidth]{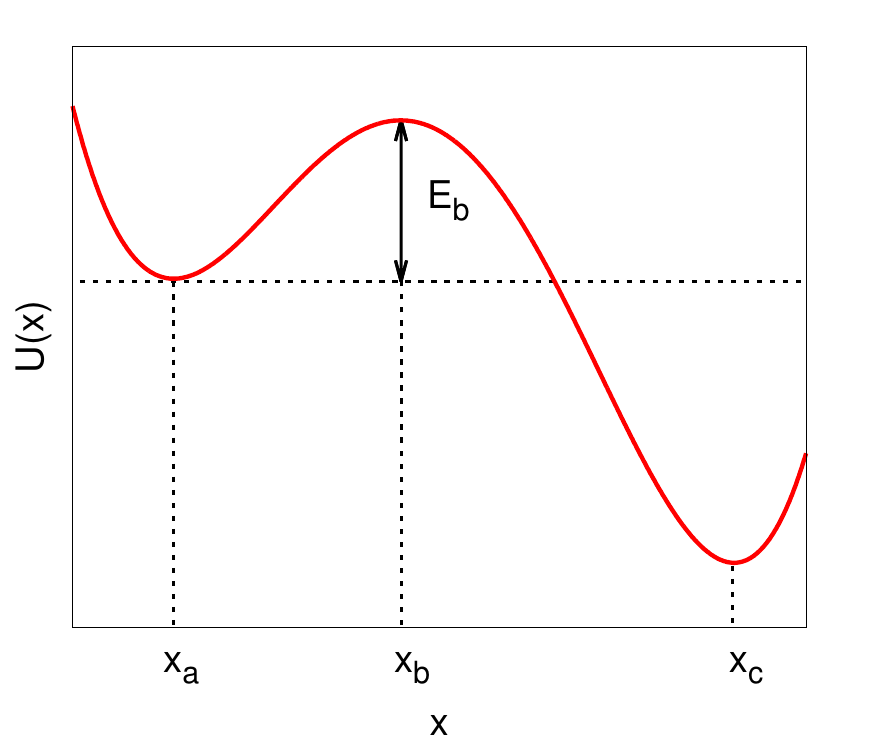}
\caption{
A schematic potential for the Kramers' model. 
For moderate to high friction, a system at a local minimum $x=x_a$ can escape the 
energy barrier ($E_b> T$) at $x=x_b$ to go to the lower bound state at $x=x_c$. 
}
\label{kramers}
\end{figure}

We consider a system initially at a local minimum $x=x_a$ (see \Fig{kramers}). 
For moderate to high friction, the system can be excited enough to overcome the 
energy barrier at $x=x_b$ and go to the lower bound state at $x=x_c$. 
Since the energy barrier is supposed to be higher than the thermal 
energy ($E_b>T$), the escape process is slow. 
It can be understood as quasistationary diffusion. 

The evolution of the system is described by 
the Fokker-Planck equation\cite{kramers}
\st
\label{kramersFP}
\left[\frac{\partial }{\partial t}
+v\frac{\partial }{\partial x}
-\mu \frac{dU(x)}{dx}\frac{\partial }{\partial v}\right]f(t,x,v)
=\gamma\frac{\partial}{\partial v}\left[vf(t,x,v)
+\mu T\frac{\partial f(t,x,v)}{\partial v}\right]
\, . 
\stp
In the diffusion limit, the system is in equilibrium everywhere 
except near the barrier at $x=x_b$. 
Thus, the dynamics may be evaluated by considering the potential around this 
point. 
In a locally harmonic approximation, the potential is given by 
$U(x)=-\omega_b^2(x-x_b)^2/2\mu+U(x_b)$, where $\omega_b$ is the angular 
frequency of the unstable state at the barrier. 
Around the barrier, there are neither sources nor sinks, and 
the distribution function satisfies the stationary Fokker-Planck equation. 

In the stationary limit, we solve the Fokker-Planck equation by using a 
Maxwellian distribution 
\st
\label{fform}
f(x,v)=\zeta(x,v) \, \exp\left[-\left(\frac{v^2}{2\mu}+U(x)\right)\Big/
T\right] \, .
\stp
With the boundary conditions $\zeta(x_a,v)=1$ and $\zeta(x_c,v)=0$, 
the Kramers' stationary solution is\cite{kramers,kramers2} 
\st
\label{kramerssol}
\zeta(x,v)=\frac{\omega_b^2}{\sqrt{2\pi \mu T\gamma \lambda}}
\int_{(x-x_b)-\lambda v/\omega_b^2}^\infty du \, 
\exp\left[-\frac{\omega_b^4u^2}{2\mu T\gamma\lambda}\right]
\, ,
\stp
where $\lambda=\sqrt{(\gamma/2)^2+\omega_b^2}-\gamma/2$. 
The stationary diffusion current is calculated by 
\st
\mathcal{J}(x_b)=\int_{-\infty}^\infty dv \, v f(x_b,v) \, ,
\stp
and the number of particles at the initial state is 
\st
\mathcal{N}(x_b)=\int_{-\infty}^{x_b}dx\int_{-\infty}^\infty dv \, f(x,v) \, ,
\stp
where we use the locally harmonic potential 
$U(x)=\omega_a^2(x-x_a)^2/2\mu+U(x_a)$ around $x=x_a$. 
The escape rate is then given by the flux over population\cite{kramers} 
\st
\label{kramersrate}
\Gamma^{kr}=\frac{\mathcal{J}(x_b)}{\mathcal{N}(x_b)}
=\frac{\lambda\omega_a}{2\pi\omega_b}
\exp\left[-E_b/T\right] \, ,
\stp
where the energy barrier is $E_b=U(x_b)-U(x_a)$. 

The Kramers' rate with the parabolic potential might give an upper 
bound of the escape rate\cite{prlett,talkner}. 
For high friction, there can be recrossing over the energy barrier, and 
anharmonic corrections of the potential barrier reduce the escape rate. 
In Ref.\cite{talkner}, the leading order effects of anharmonic 
potentials have been calculated perturbatively. 
In the next section, we use the same strategy to solve \Eq{dnseq} in a locally 
harmonic approximation.

\section{Coupling through Inverse Mass Tensors}
\label{pcncorrection}

The Kramers' quasistationary rate can be applied to two-dimensional 
potential of a dinuclear system. 
In general, the radial mode and the mass asymmetry mode are not separable, 
and they are coupled with each other. 
The coupling between two collective modes is weak for 
almost symmetric reactions, but it grows as the asymmetry 
increases\cite{microscopicmass}. 
For simplicity, we consider asymmetric dinuclear systems where the coupling 
between two modes is still weak. 
By following Ref.\cite{talkner} in which the Fokker-Planck equation is solved 
perturbatively, we investigate the 
coupling effects on the escape rates. 

Similar to \Eq{fform}, we consider a distribution function of the form 
\st
\label{fsol2d}
f(R,\eta,v_R,v_\eta)=\zeta(R,\eta,v_R,v_\eta) \, 
\exp\left[-\left(v_R^2/2\mu_R T_R+ v_\eta^2/2\mu_\eta T_\eta+
U(R,\eta)/T_{eff}\right)\right] \, .
\stp
Around the barriers $(R=R_b, \, \eta=\eta_b)$, the two-dimensional 
potential is assumed to be 
\begin{multline}
U(R,\eta)/T_{eff}=
-\omega_{R_b}^2(R-R_b)^2/2\mu_R T_R+U_{eff}(R_b)/T_R
\\
- \omega_{\eta_b}^2(\eta-\eta_b)^2/2 \mu_\eta T_\eta+
U_{eff}(\eta_b)/T_\eta \, ,
\end{multline}
where $U_{eff}(R_b)$ and $U_{eff}(\eta_b)$ are the effectively one-dimensional 
potentials in each collective coordinate. 
If there is no coupling through the inverse mass tensors, 
the escape rate is given by the Kramers' rate in \Eq{kramersrate} for 
each degree of freedom. 
In that case, we have the zeroth order equation of motion 
\st
\label{0theq}
\mathcal{L}_0 \, \zeta_0(R,\eta,v_R,v_\eta) =0 \, ,
\stp
where we have defined 
\begin{multline}
\mathcal{L}_0=
-v_R\frac{\partial}{\partial R}+
\mu_R\frac{\partial U}{\partial R}\frac{\partial}
{\partial v_R}-\gamma_R^*v_R\frac{\partial}{\partial v_R}
+\mu_R\gamma_R^*T_R\frac{\partial^2}{\partial v_R^2}
\\
-v_\eta\frac{\partial}{\partial \eta}
+\mu_\eta\frac{\partial U}{\partial \eta}
\frac{\partial}
{\partial v_\eta}-\gamma_\eta^* v_\eta\frac{\partial}{\partial v_\eta}
+\mu_\eta\gamma_\eta^* T_\eta\frac{\partial^2}{\partial v_\eta^2}
\, ,
\end{multline}
with $\gamma_i^*=\mu_i\gamma_i$ ($i=R,\eta$). 
Since $R$ and $\eta$ modes are separable at the barriers, we have 
\st
\zeta_0(R,\eta,v_R,v_\eta)=\zeta_R(R,v_R) \, 
\zeta_\eta(\eta,v_\eta) \, ,
\stp
and each factor is given by the Kramers' stationary solution in 
\Eq{kramerssol}: 
\bg
\zeta_R(R,v_R)&=&\frac{\omega_{R_b}^2}{\sqrt{2\pi \mu_R T_R\gamma_R^* \lambda_R}}
\int_{(R-R_b)-\lambda_R v_R/\omega_{R_b}^2}^\infty du \, 
\exp\left[-\frac{\omega_{R_b}^4u^2}{2\mu_R T_R\gamma_R^*\lambda_R}\right]
\, ,
\nonumber\\
\zeta_\eta(\eta,v_\eta)&=&\frac{\omega_{\eta_b}^2}{\sqrt{2\pi \mu_\eta T_\eta\gamma_\eta^* 
\lambda_\eta}}
\int_{(\eta-\eta_b)-\lambda_\eta v_\eta/\omega_{\eta_b}^2}^\infty du \, 
\exp\left[-\frac{\omega_{\eta_b}^4u^2}{2\mu_\eta T_\eta\gamma_\eta^*\lambda_\eta}\right]
\, ,
\nd
where $\lambda_i=\sqrt{(\gamma_i^*/2)^2+\omega_{i_b}^2}-\gamma_i^*/2$ 
($i=R,\eta$). 

When the relative motion and the mass asymmetry are coupled by \Eq{dnseq}, 
we define the first order operator 
\begin{multline}
\label{1stop}
\mathcal{L}_1=
\left(-\frac{\partial \mu_R}{\partial\eta}\frac{1}{\mu_R}v_R v_\eta
+\frac{1}{2} \frac{\partial\mu_\eta}{\partial R}\frac{\mu_R}{\mu_\eta^2}
  v_\eta^2 \right)\frac{\partial }{\partial v_R}
\\ 
+\left( -\frac{\partial \mu_\eta}{\partial R}\frac{1}{\mu_\eta} v_R v_\eta
+ \frac{1}{2}\frac{\partial\mu_R}{\partial\eta}\frac{\mu_\eta}{\mu_R^2} v_R^2
\right)\frac{\partial }{\partial v_\eta}
-\left(\frac{\partial\mu_\eta}{\partial R}\frac{1}{\mu_\eta} v_R
+ \frac{\partial\mu_R}{\partial\eta}\frac{1}{\mu_R} v_\eta \right) 
\, .
\end{multline}
By expanding the solution in the order of coupling, 
\st
\zeta(R,\eta,v_R,v_\eta)=\zeta_0(R,\eta,v_R,v_\eta)
+\zeta_1(R,\eta,v_R,v_\eta)+\zeta_2(R,\eta,v_R,v_\eta) + \cdots 
\, ,
\stp
we have the equations of motion\cite{talkner}: 
\bg
\label{zeta1eq}
-\mathcal{L}_0\zeta_1(R,\eta,v_R,v_\eta)
&=&\mathcal{L}_1\zeta_0(R,\eta,v_R,v_\eta) \, ,
\\
-\mathcal{L}_0\zeta_2(R,\eta,v_R,v_\eta)
&=&\mathcal{L}_1\zeta_1(R,\eta,v_R,v_\eta) \, ,
\label{zeta2eq}
\nd
and similarly for higher orders. 
In the following subsections, we will solve the first order equation of motion. 

After solving for $\zeta_1(R,\eta,v_R,v_\eta)$, we need a strategy to calculate 
the escape rate in the two-dimensional potential. 
By noting that the zeroth order result should be two independent escape 
rates, we calculate the diffusion current and the number of particles as 
follows. 
For the relative motion, the current at the barrier is 
\st
\label{jrform}
\mathcal{J}_R(R_b,\eta_b)
=
\int_{-\infty}^\infty dv_R \int_{-\infty}^\infty 
dv_\eta \, v_R f(R_b,\eta_b,v_R,v_\eta)
\, ,
\stp
and the number of particles is defined as 
\st
\label{nrform}
\mathcal{N_R}(R_b,\eta_b)=
\int_{-\infty}^{R_b}dR\int_{-\infty}^\infty dv_R 
\int_{-\infty}^\infty dv_\eta \, 
f(R,\eta_b,v_R,v_\eta) \, ,
\stp
where we have set $\eta=\eta_b$ to cancel $\exp[-U_{eff}(\eta_b)/T_\eta]$ when taking 
the flux over population\footnote{\label{nfootnote}Due to the special way 
to calculate the number of particles with two degrees of freedom, we calculate 
the number by using only the zeroth order solution.}. 
Similarly for the mass asymmetry mode, we have 
\st
\label{jetaform}
\mathcal{J}_\eta(R_b,\eta_b)
=
\int_{-\infty}^\infty dv_R \int_{-\infty}^\infty
dv_\eta \, v_\eta f(R_b,\eta_b,v_R,v_\eta)
\, ,
\stp
and
\st
\label{netaform}
\mathcal{N_\eta}(R_b,\eta_b)=
\int_{-\infty}^\infty dv_R
\int_{-\infty}^{\eta_b} d\eta \int_{-\infty}^\infty dv_\eta  \, 
f(R_b,\eta,v_R,v_\eta) \, .
\stp
Then the escape rates of two collective modes are given by the flux over 
population: 
\st
\Gamma_R=\frac{\mathcal{J}_R(R_b,\eta_b)}{\mathcal{N}_R(R_b,\eta_b)} 
\qquad \mbox{and} \qquad
\Gamma_\eta=\frac{\mathcal{J}_\eta(R_b,\eta_b)}{\mathcal{N}_\eta
(R_b,\eta_b)} \, .
\stp

Depending on reactions, not every term in \Eq{1stop} might be important to 
calculate the diffusion rates. 
To investigate the effect of each coupling, 
we solve the first order equation of motion separately. 
We present details in \Sect{nl6}, 
and only the results are presented in the following subsections.

\subsection{$\mathcal{L}_1=-\frac{\partial \mu_R}{\partial \eta}
\frac{1}{\mu_R}v_R v_\eta \frac{\partial}{\partial v_R}$}
\label{nl6}

In this subsection, we assume 
$\mathcal{L}_1=-\frac{\partial \mu_R}{\partial \eta}
\frac{1}{\mu_R}v_R v_\eta \frac{\partial}{\partial v_R}$. 
Since the right hand side of \Eq{zeta1eq} is 
\st
\mathcal{L}_1\zeta_0(R,\eta,v_R,v_\eta)=
-\frac{\partial\mu_R}{\partial\eta}\frac{1}{\mu_R} v_Rv_\eta 
\frac{\partial \zeta_R(R,v_R)}{\partial v_R} \, \zeta_\eta(\eta,v_\eta) \, ,
\stp 
we expect the first order solution to be\footnote{In \Eq{nl6sol}, there is a 
trivial term $a \, \zeta_R(R,v_R) \, \zeta_\eta(\eta,v_\eta)$, where a 
constant is chosen to be $a=0$.}
\begin{multline}
\label{nl6sol}
\zeta_1(R,\eta,v_R,v_\eta)=
\frac{\partial\mu_R}{\partial\eta}\frac{1}{\mu_R}
\left[
P_1(R,\eta,v_R,v_\eta)\frac{\partial \zeta_R(R,v_R)}{\partial v_R} 
\zeta_\eta(\eta,v_\eta)
\right.\\
\left.
+P_2(R,\eta,v_R,v_\eta)\frac{\partial \zeta_R(R,v_R)}{\partial v_R}
\frac{\partial \zeta_\eta(\eta,v_\eta)}{\partial v_\eta}\right] \, ,
\end{multline}
where $P_i(R,\eta,v_R,v_\eta)$ ($i=1, \, 2$) is a polynomial of collective 
coordinates and velocities. 
By plugging \Eq{nl6sol} into \Eq{zeta1eq}, we have 
\begin{multline}
\label{p1eq}
\left[
v_R\frac{\partial}{\partial R}-\omega_{R_b}^2(R-R_b)
\frac{\partial}{\partial v_R}
+(2\lambda_R+\gamma_R^*)v_R\frac{\partial}{\partial v_R}
-\mu_R\gamma_R^*T_R\frac{\partial^2}{\partial v_R^2}+\lambda_R
\right.
\\
\left.
+v_\eta\frac{\partial}{\partial \eta}+\omega_{\eta_b}^2(\eta-\eta_b)
\frac{\partial}
{\partial v_\eta}
+\gamma_\eta^* v_\eta\frac{\partial}{\partial v_\eta}
-\mu_\eta\gamma_\eta^* T_\eta
\frac{\partial^2}{\partial v_\eta^2} \right]P_1(R,\eta,v_R,v_\eta)
=-v_Rv_\eta \, ,
\end{multline}
and 
\begin{multline}
\label{p2eq}
\left[
v_R\frac{\partial}{\partial R}-\omega_{R_b}^2(R-R_b)
\frac{\partial}{\partial v_R}
+(2\lambda_R+\gamma_R^*)v_R\frac{\partial}{\partial v_R}
-\mu_R\gamma_R^*T_R\frac{\partial^2}{\partial v_R^2}+\lambda_R
\right.
\\
\left.
+v_\eta\frac{\partial}{\partial \eta}-\omega_{\eta_b}^2(\eta-\eta_b)
\frac{\partial}
{\partial v_\eta}
+(2\lambda_\eta+\gamma_\eta^*) v_\eta\frac{\partial}{\partial v_\eta}
-\mu_\eta\gamma_\eta^* T_\eta\frac{\partial^2}{\partial v_\eta^2}
+\lambda_\eta \right]P_2(R,\eta,v_R,v_\eta)
\\
=2\mu_\eta\gamma_\eta^* T_\eta\frac{\partial P_1(R,\eta,v_R,v_\eta)}{\partial v_\eta} \, .
\end{multline}
The polynomials are found to be 
\bg
P_1(R,\eta,v_R,v_\eta)&=&
a_1(R-R_b)(\eta-\eta_b)+a_2(R-R_b)v_\eta+a_3(\eta-\eta_b) v_R+a_4v_Rv_\eta
\, , \, \,
\nonumber\\
P_2(R,\eta,v_R,v_\eta)&=&a_5(R-R_b)+a_6v_R \, ,
\nd
where $a_i$ $(i=1, \cdots,6)$ is a constant. 
It is straightforward to determine the constants: 
\bg
\label{nl6csol}
a_1&=&\frac{\omega_{\eta_b}^2\omega_{R_b}^2(\gamma_\eta^*+\gamma_R^*+4\lambda_R)a_4}{
\lambda_R(3\lambda_\eta^2-4\lambda_R^2)+\gamma_\eta^*[\gamma_R^*(\lambda_\eta
-2\lambda_R)+\lambda_R(3\lambda_\eta-4\lambda_R)]+
\gamma_R^*(\lambda_\eta^2-2\lambda_R^2)}
\, ,
\nonumber\\
a_2&=&\frac{\omega_{R_b}^2(\gamma_R^*\lambda_R+3\lambda_R^2+\omega_{\eta_b}^2+
\omega_{R_b}^2)a_4}{
\gamma_R^*(2\lambda_R^2-\lambda_\eta^2)+\lambda_R(4\lambda_R^2-3\lambda_\eta^2)
+\gamma_\eta^*[\gamma_R^*(2\lambda_R-\lambda_\eta)+\lambda_R(4\lambda_R-
3\lambda_\eta)]}
\, ,
\nonumber\\
a_3&=&-\frac{\omega_{\eta_b}^2
[\lambda_\eta^2+\gamma_\eta^*(\lambda_\eta-\lambda_R)+\gamma_R^*\lambda_R]a_4}
{\lambda_R(3\lambda_\eta^2-4\lambda_R^2)+\gamma_\eta^*[\gamma_R^*
(\lambda_\eta-2\lambda_R)+\lambda_R(3\lambda_\eta-4\lambda_R)]+\gamma_R^*
(\lambda_\eta^2-2\lambda_R^2)
}
\, ,
\nonumber\\
a_4&=&-\frac{\lambda_\eta^2\lambda_R[\lambda_R(3\lambda_\eta^2-4\lambda_R^2)
+\gamma_\eta^*\{\gamma_R^*(\lambda_\eta-2\lambda_R)+\lambda_R(3\lambda_\eta-
4\lambda_R)\}+\gamma_R^*(\lambda_\eta^2-2\lambda_R^2)]}
{(\lambda_\eta-2\lambda_R)(2\lambda_R-\lambda_\eta+\gamma_R^*)
(2\lambda_\eta\lambda_R+\omega_{\eta_b}^2)[\lambda_R\omega_{\eta_b}^2+\lambda_\eta
(\lambda_R^2+\omega_{R_b}^2)]}
\, ,
\nonumber\\
a_5&=&\frac{2\gamma_\eta^* \mu_\eta T_\eta[(\gamma_R^*+\lambda_\eta+3\lambda_R)a_2+
\omega_{R_b}^2a_4]}
{\lambda_\eta^2+4\lambda_\eta\lambda_R+3\lambda_R^2+\gamma_R^*(\lambda_\eta+
\lambda_R)+\omega_{R_b}^2}
\, ,
\nonumber\\
a_6&=&\frac{2\gamma_\eta^* \mu_\eta T_\eta[-a_2+(\lambda_\eta+\lambda_R)a_4]}{
(\lambda_\eta+\lambda_R)(\gamma_R^*+\lambda_\eta+3\lambda_R)+\omega_{R_b}^2}
 \, .
\nd

By using Eqs.(\ref{jrform}) and (\ref{nrform}), we find 
the diffusion current and the number of particles for the relative motion to 
be 
\bg
\mathcal{J}_R(R_b,\eta_b)&=&
\sqrt{\frac{\pi\mu_\eta T_\eta}{2}}\frac{\mu_R T_R\lambda_R}{\omega_{R_b}}
\left[1
+
\frac{\partial\mu_R}{\partial\eta}\frac{1}{\mu_R}
\left(a_4 \mu_\eta T_\eta + a_6\right) 
\sqrt{\frac{2}{\pi \mu_\eta T_\eta}}
\frac{\gamma_R^*\lambda_R\lambda_\eta}{\omega_{R_b}^2\omega_{\eta_b}}
\right]
\nonumber\\
&&\qquad\times \bigg.\bigg.
\exp\left[-U_{eff}(R_b)/T_R-U_{eff}(\eta_b)/T_\eta\right]
\, , 
\nonumber\\
\mathcal{N}_R(R_b,\eta_b)&=&\sqrt{2\pi \mu_\eta T_\eta}
\frac{\pi \mu_R T_R}{\omega_{R_a}} \, 
\exp\left[-U_{eff}(R_a)/T_R-U_{eff}(\eta_b)/T_\eta\right]
\, .
\label{rnumber}
\nd
Then the flux over population is used to determine the escape rate 
\st
\label{Arate}
\Gamma_R=
\Gamma_R^{kr}\left[1+
\frac{\partial\mu_R}{\partial\eta}\frac{1}{\mu_R}
\left(a_4 \mu_\eta T_\eta + a_6\right)
\sqrt{\frac{2}{\pi\mu_\eta T_\eta}}\frac{\gamma_R^*\lambda_R\lambda_\eta}
{\omega_{R_b}^2\omega_{\eta_b}}\right] \, ,
\stp
where we have used the Kramers' escape rate in \Eq{kramersrate}. 
Similarly for the mass asymmetry mode, we have 
\bg
\mathcal{J}_\eta(R_b,\eta_b)&=&
\sqrt{\frac{\pi \mu_R T_R}{2}}\frac{\mu_\eta T_\eta\lambda_\eta}{\omega_{\eta_b}}
\,
\exp\left[-U_{eff}(R_b)/T_R-U_{eff}(\eta_b)/T_\eta\right]
\, ,
\nonumber\\
\mathcal{N}_\eta(R_b,\eta_b)&=&\sqrt{2\pi\mu_R T_R}
\frac{\pi \mu_\eta T_\eta}{\omega_{\eta_a}}
\,
\exp\left[-U_{eff}(R_b)/T_R-U_{eff}(\eta_a)/T_\eta\right]
\, ,
\label{etanumber}
\nd
and the diffusion rate is 
\st
\label{nl6Geta}
\Gamma_\eta=\Gamma_\eta^{kr} \, .
\stp
We note that the escape rate of the $R$ mode changes by the coupling, 
while the $\eta$ mode is not affected. 
The coupling effects depend on the friction coefficients and the angular 
frequencies at the barriers.

\subsection{$\mathcal{L}_1=\frac{1}{2}\frac{\partial \mu_\eta}{\partial R}
\frac{\mu_R}{\mu_\eta^2}v_\eta^2\frac{\partial}{\partial v_R}$}
\label{nl3}

The first order solution is of the form 
\begin{multline}
\zeta_1(R,\eta,v_R,v_\eta)=\frac{1}{2}
\frac{\partial\mu_\eta}{\partial R}\frac{\mu_R}{\mu_\eta^2}
\left[
P_1(R,\eta,v_R,v_\eta)\frac{\partial \zeta_R(R,v_R)}{\partial v_R} 
\zeta_\eta(\eta,v_\eta)
\right.\\
\left.
+P_2(R,\eta,v_R,v_\eta)\frac{\partial \zeta_R(R,v_R)}{\partial v_R}
\frac{\partial \zeta_\eta(\eta,v_\eta)}{\partial v_\eta}\right] \, ,
\end{multline}
and the polynomials are found to be 
\bg
P_1(R,\eta,v_R,v_\eta)&=&
b_1+b_2(\eta-\eta_b)^2+b_3(\eta-\eta_b) v_\eta+b_4 v_\eta^2 \, ,
\nonumber\\
P_2(R,\eta,v_R,v_\eta)&=&b_5(\eta-\eta_b)+b_6v_\eta \, ,
\nd
where the constants are 
\bg
\label{nl3csol}
b_1&=&\frac{2\mu_\eta \gamma_\eta^* T_\eta b_4}{\lambda_R}
\, ,
\nonumber\\
b_2&=&\frac{2\omega_{\eta_b}^4 b_4}{
(\lambda_R^2-2\lambda_\eta^2)+\gamma_\eta^*(\lambda_R-2\lambda_\eta)
} \, ,
\nonumber\\
b_3&=&-\frac{2\lambda_R \omega_{\eta_b}^2 b_4}{
(\lambda_R^2-2\lambda_\eta^2)+\gamma_\eta^*(\lambda_R-2\lambda_\eta)
} \, ,
\nonumber\\
b_4&=&\frac{
(\lambda_R^2-2\lambda_\eta^2)+\gamma_\eta^*(\lambda_R-2\lambda_\eta)
}{(\gamma_\eta^*+
\lambda_R)
(\lambda_R-2\lambda_\eta)(2\gamma_\eta^*+\lambda_R+2\lambda_\eta)
} \, ,
\nonumber\\
b_5&=&\frac{2\mu_\eta \gamma_\eta^* T_\eta[(\gamma_\eta^*+\lambda_R+3\lambda_\eta)b_3+
2\omega_{\eta_b}^2b_4]}
{\lambda_R^2+4\lambda_R\lambda_\eta+3\lambda_\eta^2+\gamma_\eta^*(\lambda_R
+\lambda_\eta)+\omega_{\eta_b}^2}
\, ,
\nonumber\\
b_6&=&\frac{2\mu_\eta\gamma_\eta^* T_\eta[-b_3+2(\lambda_R+\lambda_\eta)b_4]}
{(\lambda_R+\lambda_\eta)(\gamma_\eta^*+\lambda_R+3\lambda_\eta)
+\omega_{\eta_b}^2} \, .
\nd
The escape rates of two collective modes are 
\bg
\Gamma_R&=&\Gamma_R^{kr} \, ,
\\
\Gamma_\eta&=&\Gamma_\eta^{kr}
\left[1
+\frac{\partial\mu_\eta}{\partial R}
\frac{1}{\mu_\eta^2}
\sqrt{\frac{\mu_R}{2\pi T_R}}\frac{\lambda_R}{\omega_{R_b}}
\bigg\{b_1+b_4 \mu_\eta T_\eta\left(2+\frac{\gamma_\eta^*\lambda_\eta}{\omega_{\eta_b}^2}
\right)
+b_6\frac{\gamma_\eta^*\lambda_\eta}{\omega_{\eta_b}^2}
\bigg\}
\right] \, .
\label{Brate}
\nd
In this case, the diffusion rate of the $\eta$ mode is affected by the 
coupling.

\subsection{$\mathcal{L}_1=-\frac{\partial\mu_\eta}{\partial R}\frac{1}
{\mu_\eta}v_R v_\eta \frac{\partial}{\partial v_\eta}$}
\label{nl5}

This case is similar to \Sect{nl6} by interchanging $(R,v_R)$ and 
$(\eta,v_\eta)$. 
The first order solution has the form 
\begin{multline}
\label{nl5z1form}
\zeta_1(R,\eta,v_R,v_\eta)=
\frac{\partial\mu_\eta}{\partial R}\frac{1}{\mu_\eta}
\left[
P_1(R,\eta,v_R,v_\eta)\zeta_R(R,v_R) \frac{\partial\zeta_\eta(\eta,v_\eta)}
{\partial v_\eta}
\right.
\\
\left.+P_2(R,\eta,v_R,v_\eta)\frac{\partial\zeta_R(R,v_R)}{\partial v_R} 
\frac{\partial\zeta_\eta(\eta,v_\eta)}{\partial v_\eta}\right]
\, ,
\end{multline}
where the polynomials are 
\bg
\label{nl5p}
P_1(R,\eta,v_R,v_\eta)
&=&c_1(R-R_b)(\eta-\eta_b)
+c_2(\eta-\eta_b) v_R
+c_3(R-R_b)v_\eta
+c_4v_Rv_\eta 
\, ,
\nonumber\\
P_2(R,\eta,v_R,v_\eta)&=&c_5(\eta-\eta_b)+c_6v_\eta
\, .
\nd
The constant $c_i$ ($i=1,\cdots, 6$) is given by 
\Eq{nl6csol} except that  
$(\gamma_R^*, \, \lambda_R, \, \omega_{R_b},\mu_R,T_R)$ and 
$(\gamma_\eta^*, \, \lambda_\eta, \, \omega_{\eta_b},\mu_\eta,T_\eta)$ 
interchange. 
The escape rates of two collective modes are 
\bg
\Gamma_R&=&\Gamma_R^{kr} \, ,
\\ 
\Gamma_\eta&=&
\Gamma_\eta^{kr}
\left[1
+\frac{\partial\mu_\eta}{\partial R}\frac{1}{\mu_\eta}
\left(c_4 \mu_R T_R + c_6\right) \sqrt{\frac{2}{\pi\mu_R T_R}}
\frac{\lambda_R\lambda_\eta \gamma_\eta^*}{\omega_{R_b}\omega_{\eta_b}^2}
\right] \, .
\label{Crate}
\nd

\subsection{$\mathcal{L}_1=\frac{1}{2}\frac{\partial\mu_R}{\partial \eta}
\frac{\mu_\eta}{\mu_R^2}v_R^2\frac{\partial}{\partial v_\eta}$}
\label{nl4}

This case is similar to \Sect{nl3} by interchanging 
$(R,v_R)$ and $(\eta,v_\eta)$. 
The first order solution is of the form 
\begin{multline}
\zeta_1(R,\eta,v_R,v_\eta)=\frac{1}{2}
\frac{\partial\mu_R}{\partial \eta}\frac{\mu_\eta}{\mu_R^2}
\left[
P_1(R,\eta,v_R,v_\eta)\zeta_R(R,v_R) \frac{\partial\zeta_\eta(\eta,v_\eta)}
{\partial v_\eta}
\right.
\\
\left.+P_2(R,\eta,v_R,v_\eta)\frac{\partial\zeta_R(R,v_R)}{\partial v_R} 
\frac{\partial\zeta_\eta(\eta,v_\eta)}{\partial v_\eta}\right]
\, ,
\end{multline}
with the polynomials 
\bg
\label{nl4p}
P_1(R,\eta,v_R,v_\eta)&=&d_1+d_2(R-R_b)^2+d_3(R-R_b)v_R+d_4v_R^2 
\, ,
\nonumber\\
P_2(R,\eta,v_R,v_\eta)&=&d_5(R-R_b)+d_6v_R
\, .
\nd
The constant $d_i$ ($i=1, \cdots, 6$) is given by 
\Eq{nl3csol} except that 
$(\gamma_R^*, \, \lambda_R, \, \omega_{R_b},\mu_R,T_R)$ 
and $(\gamma_\eta^*, \, \lambda_\eta, \, \omega_{\eta_b},\mu_\eta,T_\eta)$ 
interchange. 
The escape rates of two collective modes are 
\bg
\label{Drate}
\Gamma_R&=&
\Gamma_R^{kr}\left[1+
\frac{\partial\mu_R}{\partial\eta}\frac{1}{\mu_R^2}
\sqrt{\frac{\mu_\eta}{2\pi T_\eta}}\frac{\lambda_\eta}{\omega_{\eta_b}}
\bigg\{d_1+d_4 \mu_R T_R\left(2+\frac{\gamma_R^*\lambda_R}{\omega_{R_b}^2}\right)
+d_6\frac{\gamma_R^*\lambda_R}{\omega_{R_b}^2}
\bigg\}\right] \, ,
\\
\Gamma_\eta&=&\Gamma_\eta^{kr} \, .
\nd

\subsection{$\mathcal{L}_1=-\frac{\partial\mu_\eta}{\partial R}\frac{1}
{\mu_\eta}v_R$ or 
$\mathcal{L}_1=-\frac{\partial\mu_R}{\partial\eta}\frac{1}{\mu_R}v_\eta$}
\label{nl12}

We assume $\mathcal{L}_1=-\frac{\partial\mu_\eta}{\partial R}\frac{1}
{\mu_\eta}v_R$. 
The first order solution is 
\st
\label{nl1sol}
\zeta_1(R,\eta,v_R,v_\eta)=
\frac{\partial\mu_\eta}{\partial R}\frac{1}{\mu_\eta}
P(R,\eta,v_R,v_\eta) \, 
\zeta_R(R,v_R) \, \zeta_\eta(\eta,v_\eta) \, ,
\stp
where the polynomial is found to be $P(R,\eta,v_R,v_\eta)=-(R-R_b)+c$. 
For a constant $c=0$, this coupling does not change the diffusion rates of 
two modes. 

Similarly, the coupling 
$\mathcal{L}_1=-\frac{\partial\mu_R}{\partial\eta}\frac{1}{\mu_R}v_\eta$ 
does not affect the escape rates.

\subsection{Coupling Effects on Escape Rates}
\label{exrates}

\begin{figure}
\includegraphics[width=0.45\textwidth]{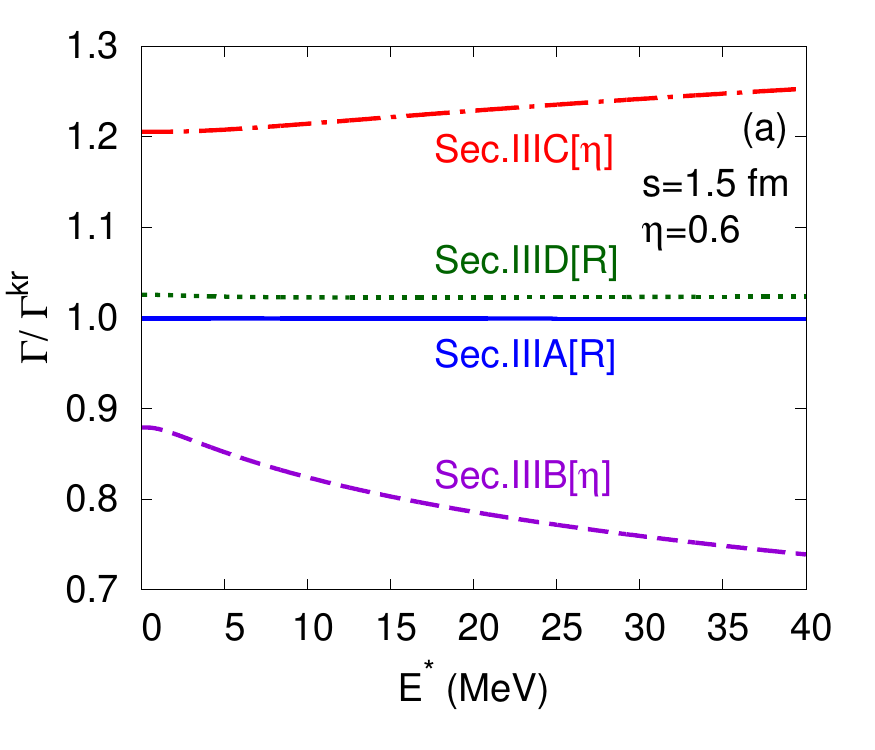}\\
\includegraphics[width=0.45\textwidth]{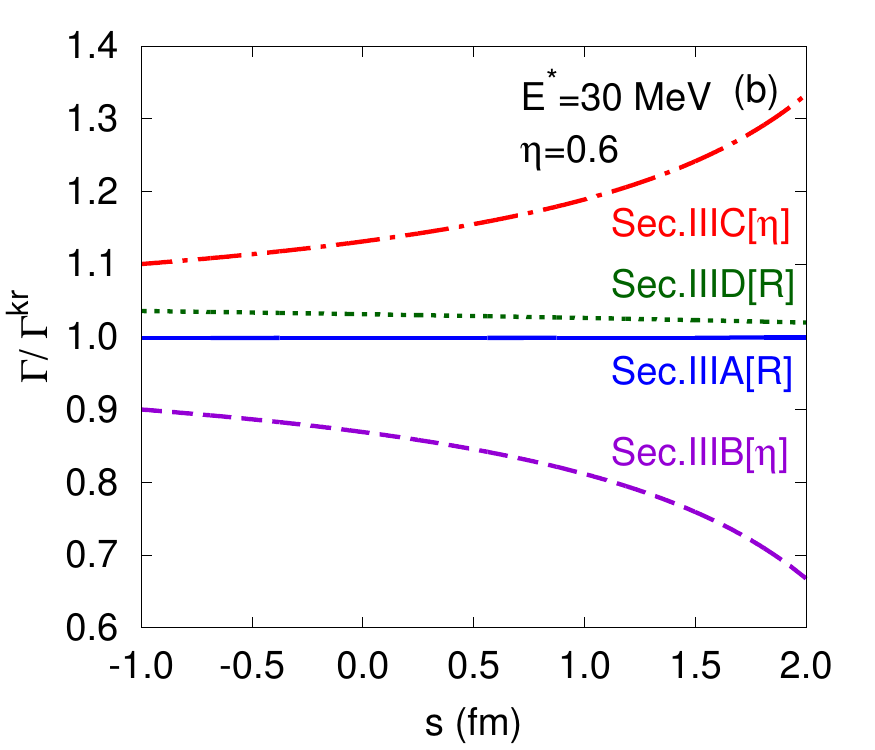}
\includegraphics[width=0.45\textwidth]{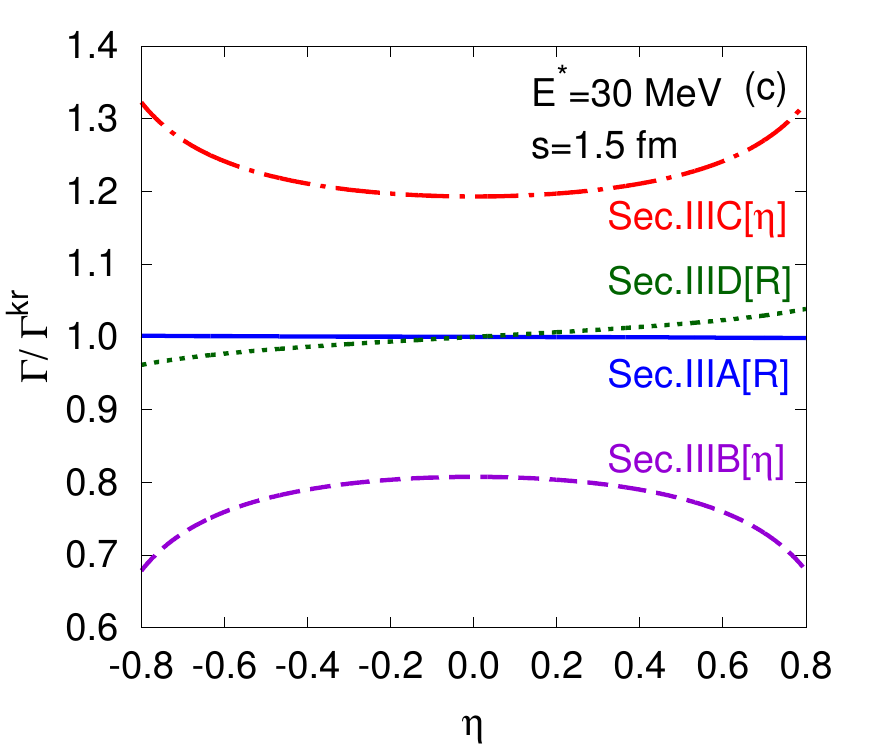}
\caption{The ratios of the escape rate to the Kramers' rate (a) as a function 
of the excitation energy for $s=1.5 \, \mbox{fm}$ and $\eta=0.6$, (b) as a 
function of the distance for $E^*=30 \, \mbox{MeV}$ and $\eta=0.6$, and (c) 
as a function 
of the mass asymmetry parameter for $E^*=30 \, \mbox{MeV}$ and $s=1.5 \, \mbox{fm}$. 
The blue solid lines are for the quasifission rate of \Eq{Arate} in \Sect{nl6}, 
the violet dashed lines are for the fusion rate of \Eq{Brate} in 
\Sect{nl3}, the red dashed-dotted lines are for the fusion rate of \Eq{Crate} in \Sect{nl5}, 
and the green dotted lines are for the quasifission rate of \Eq{Drate} in \Sect{nl4}. 
The parameters of $\hbar\omega_{R_b}=2 \, \mbox{MeV}$, 
$\hbar\omega_{\eta_b}=1 \, \mbox{MeV}$, 
$\hbar\gamma_R^*=\hbar\gamma_\eta^*=2 \, \mbox{MeV}$, 
and $A=300$ are used. 
}
\label{rate}
\end{figure}

As seen in Eqs. (\ref{Arate}), (\ref{Brate}), (\ref{Crate}), and 
(\ref{Drate}), the escape rates are affected by the coupling between two 
collective modes. 
The coupling effects generally depend on the friction coefficients 
$(\gamma_R^*,\gamma_\eta^*)$ and the angular frequencies 
$(\omega_{R_b},\omega_{\eta_b})$. 
In this subsection, we investigate the degree of the coupling effects on the 
escape rates, by using specific parameters. 
For a dinuclear system, the diffusion rates of the $R$ and $\eta$ modes 
are interpreted as the 
quasifission rate and the fusion rate, respectively (see \Sect{dnsreview}). 

The inverse mass tensors of a dinuclear system are given 
by\cite{FPkramers,microscopicmass,AA} 
\bg
\mu_R&=&\frac{4}{Am}\frac{1}{1-\eta^2}\left(1-\frac{\nu}{1-\eta^2}\right) 
\qquad \mbox{with} \qquad 
\nu=\frac{1}{A}(\xi_0-\xi_1\eta^2)(1-\xi s) \, ,
\nonumber\\
\mu_\eta&=&\frac{1}{Am}\frac{\nu}{2\sqrt{2\pi}b^2} \, , 
\nd
where $m$ is the nucleon mass, $s=R-R_1-R_2$, $\xi_0=16$, $\xi_1=17.5$, 
$\xi=0.3 \, \mbox{fm}^{-1}$, and $b=1 \, \mbox{fm}$. 
For a dinuclear system with $A=300$, we use the coefficients of 
$\hbar\omega_{R_b}=2 \, \mbox{MeV}$, $\hbar\omega_{\eta_b}=1 \, \mbox{MeV}$, 
and $\hbar\gamma_R^*=\hbar\gamma_\eta^*=2 \, \mbox{MeV}$. 

In \Fig{rate}, we present the ratios of the escape rate to the Kramers' rate. 
The blue solid lines are for the quasifission rate of \Eq{Arate} in \Sect{nl6}, 
the violet dashed lines are for the fusion rate of \Eq{Brate} in 
\Sect{nl3}, the red dashed-dotted lines are for the fusion rate of \Eq{Crate} in \Sect{nl5}, 
and the green dotted lines are for the quasifission rate of \Eq{Drate} in \Sect{nl4}. 
\Fig{rate} (a) shows the coupling effects depending on the excitation energy 
for $s=1.5 \, \mbox{fm}$ and $\eta=0.6$. 
For the particular choice of parameters, the fusion rate in \Sect{nl3} 
decreases by $10 \,$-$\, 25\%$. 
On the other hand, the coupling in \Sect{nl5} increases the fusion rate by 
$20 \,$-$\, 25\%$. 
Thus, in the presence of both couplings, there might be cancellation in some 
of the effects. 
For the quasifission rate, the coupling effects are relatively low. 
The coupling in \Sect{nl4} increases the rate by $2.5\%$, and the effect in 
\Sect{nl6} is less than $1\%$.  
\Fig{rate} (b) shows the coupling effects depending on the distance for 
$E^*=30 \, \mbox{MeV}$ and $\eta=0.6$. 
As the distance increases, the effects on the fusion rate grow due to the 
increasing mass of the $\eta$ mode. 
The fusion rates change by $10 \,$-$\, 30 \%$ while the quasifission rates 
are not affected much. 
The couplings effects also depend on the mass asymmetry as in \Fig{rate} (c). 
Dinuclear systems with higher mass asymmetry are affected more by 
the coupling through the inverse mass tensors.

\section{Dinuclear System}
\label{dnsreview}

In this section, we review the dinuclear system concept by following 
Refs.\cite{FPkramers,dnsr1,dnsr2,dns4} and apply the Kramers' escape problem to 
a dinuclear system. 
We discuss the diabatic nucleus-nucleus potential and the driving potential. 
By the Kramers' model, quasifission and fusion are described as 
diffusion in two collective coordinates. 
The cross section of evaporation residue is determined by estimating the 
fusion probability and the survival probability. 
Based on the study in the previous section, we discuss the coupling effects on 
the fusion probability and the cross section.

\subsection{Diabatic Potential}

The diabatic nucleus-nucleus potential consists of the Coulomb 
potential, the nuclear potential, and the centrifugal potential: 
\st
V(R,J)=V_C(R)+V_N(R)+V_{rot}(R,J) \, ,
\stp
where $J$ is the angular momentum. 
We consider axially symmetric systems described by 
$R_i(\theta_i)=R_{0i}\left[1+\beta_i Y_{20}(\theta_i)\right]$, 
where $\beta_i$ is the quadrupole deformation parameter and 
$R_{0i}=r_0A_i^{1/3}$ with $r_0=1.16 \, \mbox{fm}$. 
Then the Coulomb potential is given by\cite{vc} 
\st
V_C(R)=\frac{e^2Z_1Z_2}{R}
+\frac{3}{5}\frac{e^2Z_1Z_2}{R^3}\sum_{i=1,2}R_{0i}^2\beta_iY_{20}(\theta_i)
\, .
\stp
The centrifugal potential is 
\st
V_{rot}(R,J)=\frac{\hbar^2J(J+1)}{2\mathcal{I}}
\, ,
\stp
where $\mathcal{I}$ is the moment of inertia. 

The nuclear potential in the double folding form is 
\st
\label{vn}
V_N(R)=\int d\r_1\int d\r_2  \, 
\rho_1(\r_1) \, \rho_2(\bm{R}-\r_2) \,
F(\r_1-\r_2) \, ,
\stp
where $F(\r_1-\r_2)$ is the nucleon-nucleon interaction. 
A well-known ansatz for the density dependent interaction gives\cite{Migdal} 
\st
\label{int}
F(\r_1-\r_2)=C_0\left[F_{in}\frac{\rho(\r_1)}{\rho_0}+F_{ex}\left(1-\frac{
\rho(\r_1)}{\rho_0}\right)\right]\delta(\r_1-\r_2)
\, ,
\stp
where 
\st
F_{in}=f_{in}+f_{in}'\frac{N_1-Z_1}{A_1}\frac{N_2-Z_2}{A_2} \, ,
\stp
and similarly for $F_{ex}$. 
The density $\rho(\r)=\rho_1(\r)+\rho_2(\r)$ is given by the 
Wood-Saxon form
\st
\rho_i(\r)=\frac{\rho_0}{1+\exp[(r-R_i(\theta_i))/a_0]} \, . 
\stp
We use the parameters of
$C_0=300 \, \mbox{MeV}\, \mbox{fm}^3$, 
$f_{in}=0.09$, $f_{in}'=0.42$, $f_{ex}=-2.59$, $f_{ex}'=0.54$, 
$\rho_0=0.17\, \mbox{fm}^{-3}$, and $a_0=0.55 \, \mbox{fm}$. 

\begin{figure}
\includegraphics[width=0.45\textwidth]{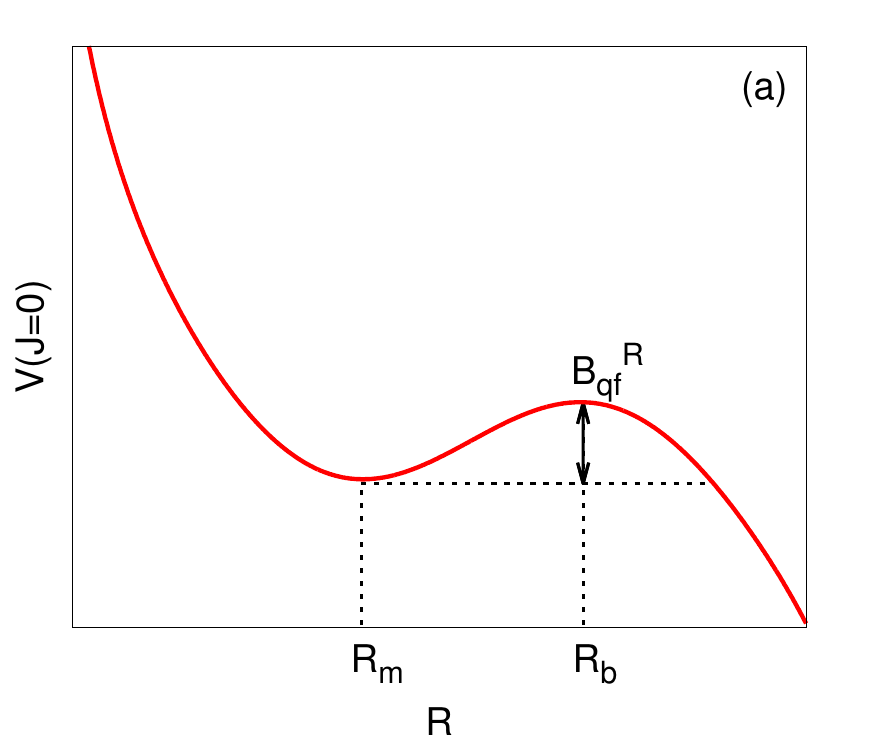}
\includegraphics[width=0.45\textwidth]{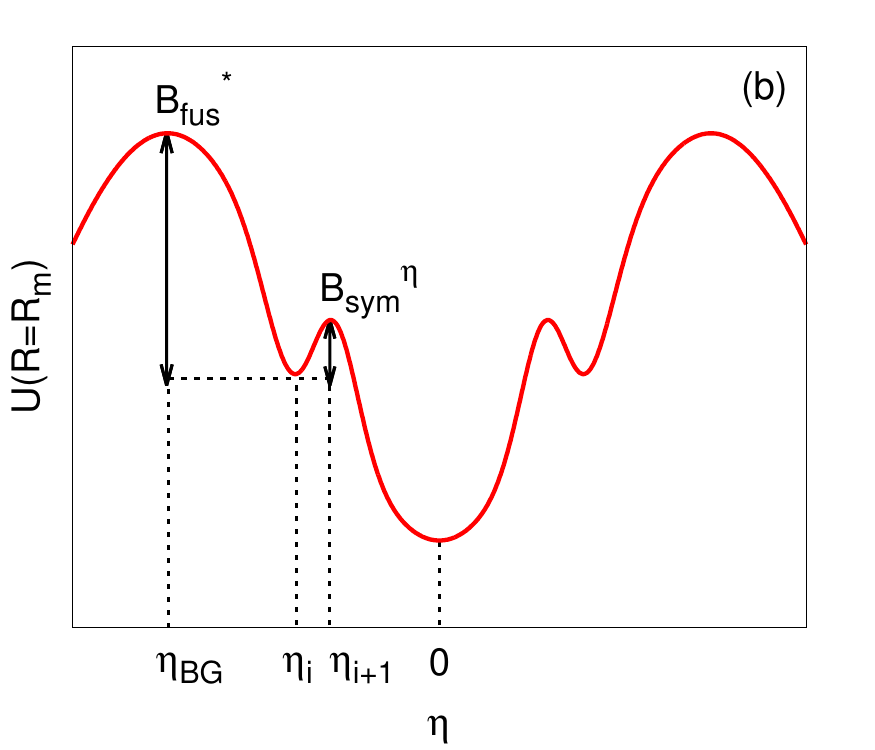}
\caption{(a) The diabatic nucleus-nucleus potential has an energy 
barrier, $B_{qf}^R=V(R_b)-V(R_m)$. 
(b) The driving potential has the inner fusion barrier, 
$B_{fus}^*=U(\eta_{BG})-U(\eta_i)$. 
There is another barrier for symmetrization of 
a dinuclear system, 
$B_{sym}^\eta=U(\eta_{i+1})-U(\eta_i)$. 
The quasifission barrier is defined as 
$B_{qf}=\mbox{Min}[B_{qf}^R,B_{sym}^\eta]$. 
}
\label{sch}
\end{figure}

For $J=0$, a schematic nucleus-nucleus potential is shown in \Fig{sch} (a). 
In contrast to the adiabatic potential, the diabatic nucleus-nucleus potential 
diverges at short distances. 
It has an energy barrier, $B_{qf}^R=V(R_b)-V(R_m)$. 
The potential energy of a dinuclear system is 
\st
U(R,\eta,J)=
B_1+B_2+V(R,J) \, ,
\stp
where $B_i$ ($i=1, \, 2$) is the binding energy including shell effects. 
A schematic of the driving potential is shown in \Fig{sch} (b). 
It has the inner fusion barrier, 
$B_{fus}^*=U(\eta_{BG})-U(\eta_i)$, 
which is the difference between the driving potential 
at the Businaro-Gallone point $\eta=\eta_{BG}$ and the initial value at 
$\eta=\eta_i$. 
$B_{sym}^\eta=U(\eta_{i+1})-U(\eta_i)$ is the barrier 
for symmetrization of a dinuclear system. 
The quasifission barrier is defined as 
$B_{qf}=\mbox{Min}[B_{qf}^R,B_{sym}^\eta]$ \footnote{Unlike the assumption made 
in \Sect{pcncorrection}, $R$ and $\eta$ modes are 
coupled with each other through the potential even if there is no 
coupling through the inverse mass tensors.}. 
 
In the dinuclear system concept, fusion reaction can be described as follows. 
The initial dinuclear system is supposed to be at the bottom of the potential 
pocket since the system at the minimum survives longest. 
If the system overcomes the quasifission barrier, it decays. 
The escape over the barrier can be understood as diffusion in the relative 
distance. 
On the other hand, if a dinuclear system overcomes the inner fusion barrier, it undergoes 
fusion and forms the compound nucleus. 
Fusion can be understood as diffusion in the mass asymmetry parameter through 
transfer of nucleons. 
There is competition between quasifission and fusion. 
The fusion probability can be estimated by the Kramers' model. 
After fusion, the system is highly excited and is likely to fission. 
However, there is a possibility that a dinuclear system remains as 
evaporation residue by emitting neutrons. 
This possibility is defined as the survival probability. 
In the following subsections, we discuss how to calculate the cross section of 
evaporation residue by estimating the fusion probability and 
the survival probability.

\subsection{Evaporation Residue}

The cross section of evaporation residue is determined by the effective capture 
cross section, the fusion probability, and the survival 
probability\cite{zubov}: 
\st
\label{sereff}
\sigma_{ER}(E_{cm})= \sigma_{cap}(E_{cm}) \, P_{CN}(E_{cm}) \, 
W_{sur}(E_{cm}) \, ,
\stp
with 
\st
\sigma_{cap}(E_{cm})=\frac{\pi \hbar^2}{2\mu E_{cm}}(J_{max}+1)^2 \, 
\mathcal{T}(E_{cm}) \, ,
\stp
where $\mu$ is the reduced mass, $E_{cm}$ is the bombarding energy in the 
center of mass system, and $\mathcal{T}(E_{cm})$ is the transmission 
probability. 

The fusion probability describes the possibility for a dinuclear system 
to overcome the inner fusion barrier, competing with quasifission. 
We apply the Kramers' escape problem to a dinuclear system with 
the quasifission barrier, $B_{qf}=U_{eff}(R_b)-U_{eff}(R_a)$, and the inner fusion 
barrier, $B_{fus}^*=U_{eff}(\eta_b)-U_{eff}(\eta_a)$. 
Then the fusion probability can be estimated 
as\footnote{$\Gamma_R+\Gamma_\eta$ is 
the sum of the diffusion rates of 
$R$ and $\eta$ including symmetrization of a dinuclear system.}\cite{dns4,dns3,dnsr1} 
\st
\label{pcndef}
P_{CN}=\frac{\Gamma_\eta}{\Gamma_R+\Gamma_\eta} \, . 
\stp
As discussed in \Sect{pcncorrection}, the coupling effects on the escape rates 
depend on the friction coefficients and the angular frequencies at the 
barriers.  
For the particular choice of parameters in \Sect{exrates}, the coupling 
in \Sect{nl3} reduces the fusion rate, so we expect the fusion probability 
to decrease. 
On the other hand, the coupling in \Sect{nl5} might increase the fusion 
probability. 
The effects on the quasifission rate in \Sect{nl6} and \Sect{nl4} might be 
ignored, in comparison to the fusion rate. 

Although the values of the coefficients are not known, 
a phenomenological formula for the fusion probability 
has been obtained\cite{zubov}: 
\st
\label{pcnform}
P_{CN}=
\frac{1.25 \, \exp\left[-(B_{fus}^*-B_{qf})/T_{DNS}\right]}
{1+1.25 \, \exp\left[-(B_{fus}^*-B_{qf})/T_{DNS}\right]} \, ,
\stp
where the temperature of a dinuclear system is 
$T_{DNS}=\sqrt{E_{DNS}^*/a}$ with the excitation energy, 
$E_{DNS}^*=E_{CN}^*-U(\eta_i)$. 
Based on the study in \Sect{pcncorrection}, 
the coupling effects might be included in the coefficient of the exponential 
term in \Eq{pcnform}. 
Thus, the variation of this coefficient allows us to estimate the 
degree of the coupling effects. 

After fusion, the compound nucleus can remain as evaporation residue 
by emitting neutrons, not to fission. 
This possibility is described by the survival probability. 
The survival probability can be estimated to be\footnote{Except fission and 
neutron emission, we ignore all other processes such as 
$\alpha$ particle emission and $\gamma$ ray emission.}\cite{wsur,wsur2} 
\st
W_{sur}(E_{CN}^*)
= P_{xn}(E_{CN}^* )\, 
\prod_{i=1}^x\frac{\Gamma_{n}(E_{CN,i}^*)}{\Gamma_{f}(E_{CN,i}^*)
+\Gamma_{n}(E_{CN,i}^*)} \, ,
\stp
where $P_{xn}(E_{CN}^*)$ is the probability of emitting $x$ neutrons 
at the excitation energy $E_{CN}^*$, 
$\Gamma_f(E_{CN,i}^*)$ is the width of 
fission at the excitation energy $E_{CN,i}^*$, 
and $\Gamma_n(E_{CN,i}^*)$ is the width of neutron emission at $E_{CN,i}^*$. 
Since an emitted neutron carries the average energy of $B_n+2T$ ($B_n$ is the 
neutron separation energy), we have 
$E_{CN,i+1}^*=E_{CN,i}^*-(B_{n,i}+2T_{i})$, where $E_{CN,1}^*=E_{CN}^*$. 

For $x\ge 2$, the probability of $x$ neutron evaporation is 
\st
\label{Pxn}
P_{xn}(E_{CN}^*)=P[x]-P[x+1] \, ,
\stp
where 
\st
P[x]\equiv 1-e^{-\Delta_x/T}\left(1+\sum_{i=1}^{2x-3}\frac{(\Delta_x/T)^i}{i!}\right)
\qquad
\mbox{with}
\qquad
\Delta_x=E_{CN}^*-\sum_{i=1}^xB_{n,i} \, ,
\stp
and the effective temperature is $T=\sqrt{E_{CN}^*/1.5 \, a}$. 
For $1$n channel, we use 
\st
P_{1n}(E_{CN}^*)=\exp\left[-(E_{CN}^*-B_n-2T)^2/2\sigma^2\right] \, ,
\stp
where $T=\sqrt{E_{CN}^*/a}$ and $\sigma=2.5 \, \mbox{MeV}$. 
The ratio of the neutron emission width to the fission width is 
given by\cite{VH,dns4} 
\st
\label{ratio}
\frac{\Gamma_n}{\Gamma_f}=
\frac{4A^{2/3}(E_{CN}^*-B_n)}{k[2\sqrt{a(E_{CN}^*-B_f)}-1]}
\exp\left[2\sqrt{a}\left(\sqrt{E_{CN}^*-B_n}-\sqrt{E_{CN}^*-B_f}\right)
\right] \, .
\stp
Here, $k=9.8 \, \mbox{MeV}$, and the fission barrier depends on the excitation energy as 
$B_f(E_{CN}^*)=B_f(E_{CN}^*=0) \, \exp\left[-E_{CN}^*/E_d\right]$ with 
$E_d=0.4 \, A^{4/3}/a$\cite{edref}.

\subsection{Coupling Effects}
\label{exreactions}

In this section, we investigate the degree of the coupling effects on the 
fusion probability and the cross section for the reaction 
$_{26}^{58}\mbox{Fe}+_{\, \, \, 82}^{208}\mbox{Pb}
\rightarrow _{\, \, \, \, \, \, \, \, 108}^{266-x}\mbox{Hs}+x\mbox{n}$ 
$(x=1, 2)$. 
We use $\theta_1=0$, $\theta_2=\pi$, $J_{max}=10$, and 
$\mathcal{T}(E_{cm})=1$ for estimation. 
The nuclear data such as the Q value, the fission barrier, and the neutron 
separation energy is from Refs.\cite{data1,data2}. 

\begin{figure}
\includegraphics[width=0.45\textwidth]{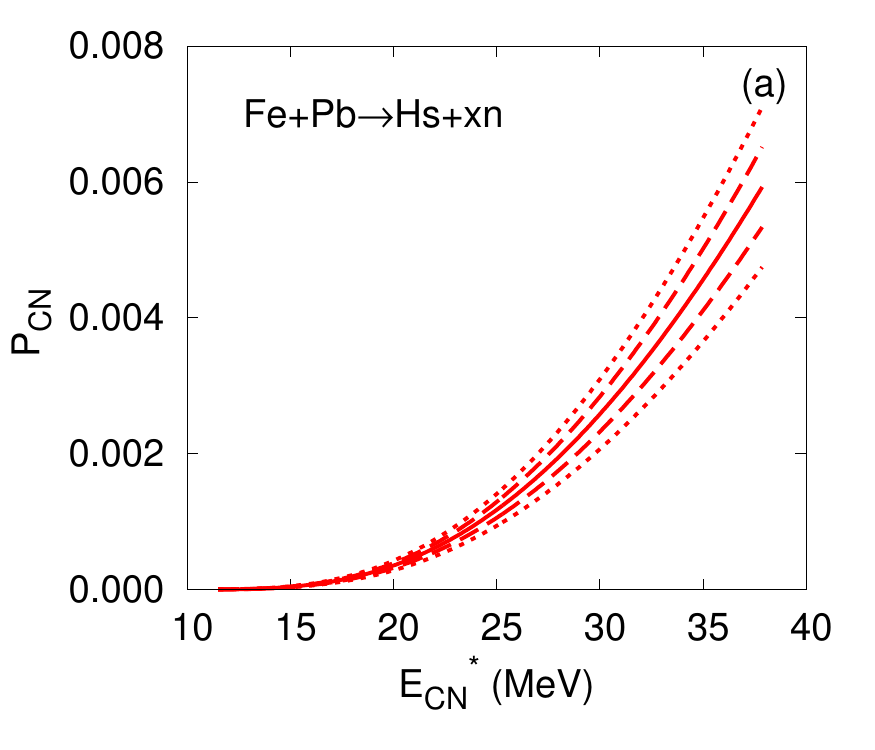}
\includegraphics[width=0.45\textwidth]{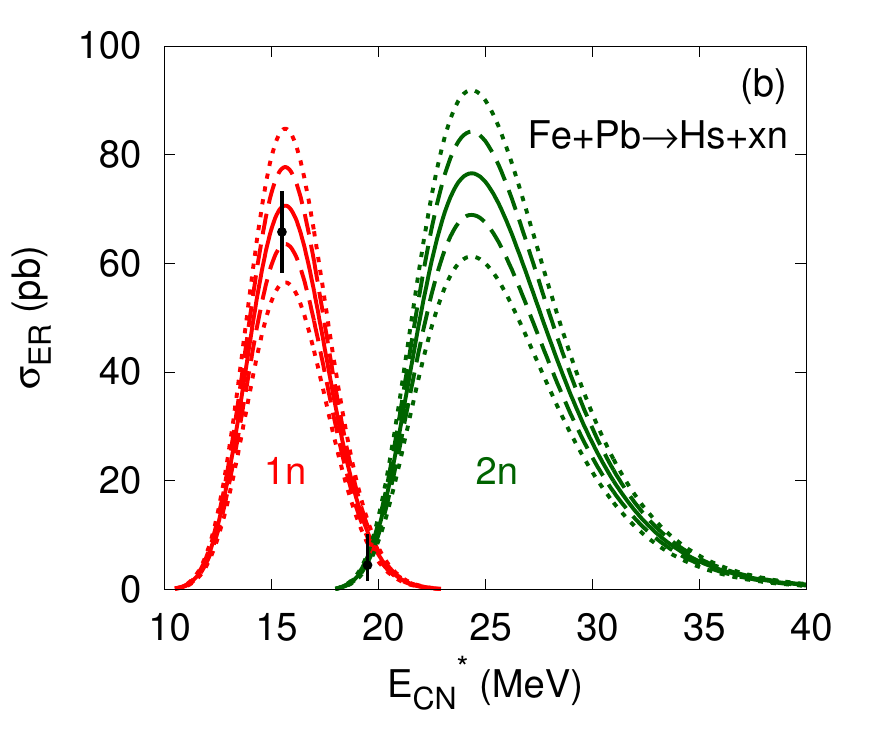}
\caption{
(a) The fusion probability for the reaction 
$_{26}^{58}\mbox{Fe}+_{\, \, \, 82}^{208}\mbox{Pb}
\rightarrow _{\, \, \, \, \, \, \, \, 108}^{266-x}\mbox{Hs}+x\mbox{n}$. 
The solid line represents the phenomenological formula given by \Eq{pcnform}, 
the dashed lines include $10\%$ coupling effects compared to the phenomenological formula, 
and the dotted lines include $20\%$ coupling effects. 
(b) The cross section of the evaporation residue for the reaction
$_{26}^{58}\mbox{Fe}+_{\, \, \, 82}^{208}\mbox{Pb}
\rightarrow _{\, \, \, \, \, \, \, \, 108}^{266-x}\mbox{Hs}+x\mbox{n}$. 
The red lines are for the $1\mbox{n}$ channel, and the green lines are for 
the $2\mbox{n}$ channel. 
The solid lines are calculated with the fusion probability given by 
\Eq{pcnform}, 
the dashed lines include $10\%$ coupling effects on the fusion 
probability, and the dotted lines include $20\%$ coupling effects. 
The black points represent the experimental data with 
errors\cite{dns3}. 
}
\label{FePbfig}
\end{figure}

\Fig{FePbfig}(a) shows the fusion probability depending on the excitation 
energy. 
The solid line is calculated with the phenomenological formula in \Eq{pcnform}, 
the dashed lines include $10\%$ coupling effects compared to the 
phenomenological formula, and the dotted lines include $20\%$ coupling 
effects. 
We mention that $10 \,$-$ \, 20\%$ effects are caused by corrections to 
both the quasifission rate and the fusion rate. 
The fusion probability generally grows as the excitation energy increases. 
The degree of increase depends on the degree of couplings. 
\Fig{FePbfig}(b) shows the cross section of evaporation residue calculated by 
\Eq{sereff}. 
The red lines are for the $1$n evaporation channel, and the green lines are 
for the $2$n channel. 
The solid lines are calculated with the fusion probability given by the 
phenomenological formula. 
The dashed and dotted lines are calculated with the fusion probability 
including $10\%$ and $20\%$ coupling effects, respectively. 
As the excitation energy increases, the cross section initially grows and then 
decreases due to the survival probability. 
At the maximum of the cross section, the coupling effect is largest. 
The black points show the experimental data with errors\cite{dns3}. 
The cross section calculated with the phenomenological fusion probability 
agrees with the experimental data within the errors. 
However, the coupling effects might produce uncertainties on the cross 
section. 
We note that the magnitude of the error is comparable to $10\,$-$\, 20\%$ 
coupling effects at the maximum of the cross section. 

We should mention that there are other factors contributing to the 
uncertainties of the cross section, in addition to the coupling effects between 
two collective modes of a dinuclear system. 
Especially, the degree of the survival probability highly depends on 
calculation methods and parameter values. 
However, in \Fig{FePbfig} we present the degree of the coupling effects for a 
reaction in which the calculated cross section agrees with the experimental 
data.

\section{Summary and Discussions}
\label{sumdiscuss}

In this work, we considered the Kramers' escape problem with two collective 
modes and 
calculated the diffusion rates in the weak coupling limit. 
Due to the coupling through the inverse mass tensors, the rates have 
corrections depending on the friction coefficients and the angular 
frequencies at the barriers. 
Our results are given by 
Eqs.(\ref{Arate}), (\ref{Brate}), (\ref{Crate}), and (\ref{Drate}). 
They can be generally applied to any escape problem in a quantum 
mechanical system with friction and diffusion. 

We have applied the escape problem to a dinuclear system 
to estimate the fusion probability. 
For the coefficients of 
$\hbar\omega_{R_b}=2 \, \mbox{MeV}$, $\hbar\omega_{\eta_b}=1 \, \mbox{MeV}$, 
and $\hbar\gamma_R^*=\hbar\gamma_\eta^*=2 \, \mbox{MeV}$, 
we present the coupling effects on the diffusion rates in \Fig{rate}. 
The fusion rate changes up to $30\%$ while the quasifission rate is barely 
affected by comparison. 
Since the fusion probability is determined by the diffusion rates of two 
collective modes, the couplings affect the fusion probability 
and the cross section of evaporation residue. 
\Fig{FePbfig} shows $10\, $-$\, 20\%$ coupling effects for the reaction 
$_{26}^{58}\mbox{Fe}+_{\, \, \, 82}^{208}\mbox{Pb}
\rightarrow _{\, \, \, \, \, \, \, \, 108}^{266-x}\mbox{Hs}+x\mbox{n}$ 
($x=1,2$). 
The effects depend on the excitation energy, and they are the largest at the 
maximum of the cross section. 
While the calculated cross section agrees with data, 
the experimental uncertainties are comparable to $10 \,$-$\, 20\%$ 
coupling effects at the maximum of the cross section. 

We should mention that there are other uncertainties in estimating the cross 
section of evaporation residue. 
Some of the effects might be greater than the coupling effects on the fusion 
probability. 
Especially, the survival probability highly depends on the calculation 
methods and parameter values. 
However, the purpose of this work is to investigate the coupling effects on 
the fusion probability for more quantitative analysis of data. 
We calculated the leading order corrections to the quasifission rate and the 
fusion rate. 
The effects of each coupling might be important to understand fusion 
depending on reactions. 
Since we do not have complete understanding of fusion dynamics nor 
nuclear data in details, we need a phenomenological strategy to compare the 
analytical results with experiments. 

We focused on asymmetric dinuclear systems in the presence of the coupling 
through the inverse mass tensors. 
We need many assumptions to obtain \Eq{dnseq}, but the general description 
of a dinuclear system involves other effects. 
First, we ignored $\partial \mu_R/\partial R$ and 
$\partial \mu_\eta/\partial \eta$ which might be important in some reactions.  
Second, we have non-diagonal components of the inverse mass tensor, the 
friction tensor, and the diffusion tensor as well as the diagonal components 
of the kinetic energy. 
By including those terms, the coupling between the relative motion and the mass 
asymmetry might be related to the neck dynamics. 
Third, we used a locally harmonic approximation, but the diabatic potential 
might need anharmonic corrections. 
Finally, the Fokker-Planck equation can be extended to include the second 
derivatives with respect to collective coordinates (and with respect to a 
collective coordinate and the corresponding conjugate momentum)\cite{Drr}. 
In principle, all the above effects can be calculated in a similar way 
used in this work. 
We hope to discuss the general analysis in future presentations.

\bigskip
\bigskip

\noindent{\bf ACKNOWLEDGMENTS}

We would like to thank G. G. Adamian and N. V. Antonenko for useful discussions 
and for providing their numerical code to calculate the diabatic potential. 
This work is supported by the Rare Isotope Science Project of Institute for 
Basic Science funded by Ministry of Science, ICT and Future Planning and 
National Research Foundation of Korea (2013M7A1A1075766).

\bibliography{fusion}

\end{document}